# QUANTITATIVE PERFUSION MAPS USING A NOVELTY SPATIOTEMPORAL CONVOLUTIONAL NEURAL NETWORK

*Anbo Cao[1,2], Pin-Yu Le[3], Zhonghui Qie[5], Haseeb Hassan[2], Yingwei Guo[6], Asim Zaman[2,7], Jiaxi Lu[1,2], Xueqiang Zeng[1,2], Huihui Yang[1,2], Xiaoqiang Miao[2,6], Taiyu Han[1,2], Guangtao Huang[1,2], Yan Kang[1,2,6,8\*], Yu Luo[9\*] and Jia Guo[4\*]*

[1]School of Applied Technology, Shenzhen University, Shenzhen 518060, China;
[2]College of Health Science and Environmental Engineering, Shenzhen Technology University, Shenzhen 518118, China;
[3]Department of Biomedical Engineering, Columbia University, New York, NY 10027, USA;
[4]Department of Psychiatry, Columbia University, New York, NY 10027, USA;
[5]Zuckerman Institute, Columbia University, New York, NY 10027, USA;
[6]College of Medicine and Biological Information Engineering, Northeastern University, Shenyang 110169, China;
[7]School of Biomedical Engineering, Medical School, Shenzhen University, Shenzhen 518060, China;
[8]Engineering Research Centre of Medical Imaging and Intelligent Analysis, Ministry of Education, Shenyang 110169, China;
[9]Shanghai Fourth People's Hospital Affiliated to Tongji University School of Medicine, Shanghai 200434, China

**ABSTRACT**

Dynamic susceptibility contrast magnetic resonance imaging (DSC-MRI) is widely used to evaluate acute ischemic stroke to distinguish salvageable tissue and infarct core. For this purpose, traditional methods employ deconvolution techniques, like singular value decomposition, which are known to be vulnerable to noise, potentially distorting the derived perfusion parameters. However, deep learning technology could leverage it, which can accurately estimate clinical perfusion parameters compared to traditional clinical approaches. Therefore, this study presents a perfusion parameters estimation network that considers spatial and temporal information, the Spatiotemporal Network (ST-Net), for the first time. The proposed network comprises a designed physical loss function to enhance model performance further. The results indicate that the network can accurately estimate perfusion parameters, including cerebral blood volume (CBV), cerebral blood flow (CBF), and time to maximum of the residual function (Tmax). The structural similarity index (SSIM) mean values for CBV, CBF, and Tmax parameters were 0.952, 0.943, and 0.863, respectively. The DICE score for the hypo-perfused region reached 0.859, demonstrating high consistency. The proposed model also maintains time efficiency, closely approaching the performance of commercial gold-standard software.

## 1. INTRODUCTION

Acute Ischemic Stroke (AIS) is the most common type of stroke. The concept of "time is brain" underscores the urgency of restoring blocked blood vessels promptly to enhance patient recovery. Dynamic susceptibility contrast magnetic resonance imaging (DSC-MRI) is used to quantitatively estimate cerebral perfusion parameters, aiding in the assessment of irreversibly infarcted brain tissue (infarct core) and potentially salvageable hypo-perfused brain tissue (penumbra) [1]. This information is crucial for evaluating the potential benefits of vascular recanalization intervention therapy and enhancing clinical treatment outcomes. These physiological parameters can be derived using the indicator dilution theory [2] and deconvolution algorithms, as described in the following equations.

$$\Delta R_2^*(t) = -\frac{1}{TE}\log(\frac{S(t)}{S_0}) \quad (1)$$

$$c_t(t) = x_1 \Delta R_2^*(t) \quad (2)$$

$$r(t) = c_t(t) \otimes^{-1} c_a(t) \quad (3)$$

$$CBV = 100 \cdot \frac{k_{AV}}{\rho} \frac{(1-H_{SV})}{(1-H_{LV})} \frac{\int c_t(t)dt}{\int c_a(t)dt}, [ml/100g] \quad (4)$$

$$CBF = 100 \cdot 60 \cdot \frac{k_{AV}}{\rho} \frac{(1-H_{SV})}{(1-H_{LV})} [\max(r(t))], [ml/100g/\min] \quad (5)$$

$$MTT = \frac{60 \cdot CBV}{CBF}, [s] \quad (6)$$

$$T\max = \arg\max_t [r(t)], [s] \quad (7)$$

$$k_{AV} = \frac{\int c_a(t)dt}{\int c_v(t)dt} \quad (8)$$

Where S(t) represents the MR signal change over time, with $S_0$ as the baseline MR signal before the contrast agent arrival. TE is the echo time, $c_t(t)$ is the contrast concentration change in the tissue, $c_a(t)$ in arterial blood vessels, and $c_v(t)$ in venous blood vessels. $H_{LV}$ and $H_{SV}$ represent the correction for different hematocrit levels in large vessels and capillaries. ρ represents constant brain density and the symbol $\otimes^{-1}$ denotes the deconvolution operation.

However, the traditional deconvolution algorithms [3-5] are sensitive to noise in time-concentration curves, leading to distortions and impacting the practical use of perfusion parameters.

While previous studies [6-7] have proposed noise-handling methods, they often require significant computation time.

In contrast, the rise of deep learning technology has sparked interest in using Convolutional Neural Networks (CNNs) for MRI data analysis. CNNs have proven to accurately predict MRI parameters [8-9] and can learn noise characteristics [10], extracting valuable information from the data. Our research introduces a novel CNN model, the Spatiotemporal Network (ST-Net), designed for quantifying perfusion parameters such as CBV, CBF, Tmax, and MTT = CBV/CBF. ST-Net leverages spatial and temporal image information, demonstrating high consistency with clinical gold standards in predicting parameters and proficiently identifying hypo-perfused brain regions using Tmax.

## 2. MATERIALS AND METHODS

### 2.1. Data and preprocessing

The gold standard perfusion parameter maps were generated using RAPID [11] software. Arterial Input Function (AIF) and Venous Output Function (VOF) regions were delineated by two experienced clinical physicians, totaling 138 patient data. The raw images, with dimensions of 256x256x20x50 representing rows, columns, slices, and acquisitions, respectively, were converted from DICOM to NIFTI format. FreeSurfer [12] was used for within-subject robust registration, and brain tissue and hypo-perfused regions masks were determined based on the method in RAPID [11].

### 2.2. Network architecture

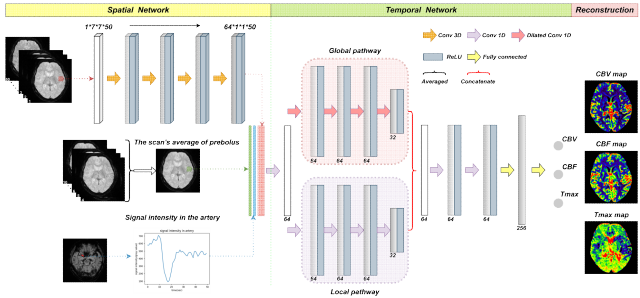

**Fig. 1. ST-Net architecture.** The four-dimensional DSC-MRI scans were first cropped to 7x7x50 patches. We then extracted spatial information using a three-dimensional convolutional neural network (CNN) encoder. Concatenation of spatial features with two additional channels: (1) The average of DSC-MRI signal before contrast agent injection for each voxel and (2) signal intensity change in arterials. The size of the output features for each layer is provided in the figure. A Leaky ReLU activation follows each fully connected layer. The outputs from the proposed ST-Net are CBV, CBF, and Tmax, which are reconstructed to acquire whole-brain parameter maps.

*2.2.1. Spatial network*
All voxels within the brain tissue region are cropped into patches of size 7x7x50 and fed into a 3D CNN encoder to extract the spatial features and preserve the time dimension. The channel progression is 1, 8, 16, 32, 64. The initial convolutional layer uses a (3, 3, 1) kernel with (1, 1, 0) padding and a stride of 1. Subsequent layers employ a (3, 3, 1) kernel with a stride of 1 and no padding. A ReLU activation follows each convolutional layer. The final output is 64x1x1x50, compressed to 64x50, and serves as input for the temporal network.

*2.2.2. Temporal network*
The input (66x50) is based on the output of the spatial network, i.e., (64x50) and two additional channels. The first channel represents the baseline signal intensity of voxels before the contrast agent arrives, spreading to the same length as the time dimension (50). The second channel denotes MR signal intensity changes in the arterial vessels, manually labeled by clinical doctors. We first used a 1D CNN layer to merge the spatial information with the additional channels. Subsequently, the input is divided into two parallel CNN pathways: The global pathway for extracting long-term temporal features and the local pathway for extracting short-term temporal features. Finally, using two 1D CNN layers and two fully connected layers, the features extracted from global and local pathways are integrated to predict the three parameter values (CBV, CBF, Tmax) corresponding to each voxel. It is worth noting that since MTT can be calculated from CBV and CBF, the model output does not include it. We added dropout layers after the fully connected layers to prevent overfitting. The ST-Net architecture is illustrated in Fig. 1.

*2.2.3. Loss function*
The loss function has two components. First, the supervised loss function uses the mean absolute error (MAE) between the model's predicted and the gold standard parameters (CBV, CBF, and Tmax). Second, the unsupervised physical loss is based on mathematical theory (equation (1,2,4,8)) and uses the MAE between the integral of the time-concentration function obtained from the raw images and the predicted CBV parameters, as shown in equation (9). During training, we adjusted the weights of these two loss functions to optimize the model.

$$Loss_{network} = L1loss(GT_{(cbv,cbf,t\max)}, \Pred_{(cbv,cbf,t\max)})$$
$$+ w * L1loss(\int c_t(t)dt, f(\Pred_{cbv})) \quad (9)$$

Where GT are the gold standard parameters, Pred are the predicted parameters. w is the weight of the physical loss. $f(Pred_{cbv})$ is the integral of the time-concentration function obtained from the predicted CBV parameter based on equation (4).

*2.2.4. Model's training hyperparameters*
The training of the ST-Net model uses Adam as the optimizer [13], with training epochs 400, batch size 512, and learning rate 1e-4. The training is stopped early if the validation loss does not decrease for 50 consecutive epochs during the training process. The model is trained using NVIDIA A100 GPU and the PyTorch deep learning framework. Our code can be found at https://github.com/muccaoanbo/Code_ST-Net.

### 2.3. Data partitioning and training strategies

Given the preprocessing, seventy patient data had hypo-perfused area labels. Of these, 52 were randomly chosen for the training and validation. The remaining 86 of 138 were for testing, including 18 with hypo-perfused area labels.

During training, two types of voxels from regions of interest (ROI) were used as inputs: normal brain tissue and hypo-perfused area, with the ratio dynamically adjusted to address potential data imbalance issues. We also noticed the stride between patches. Therefore, two other hyperparameters such as ratio and stride were set, and the model's performance was compared with and without the physical loss function.

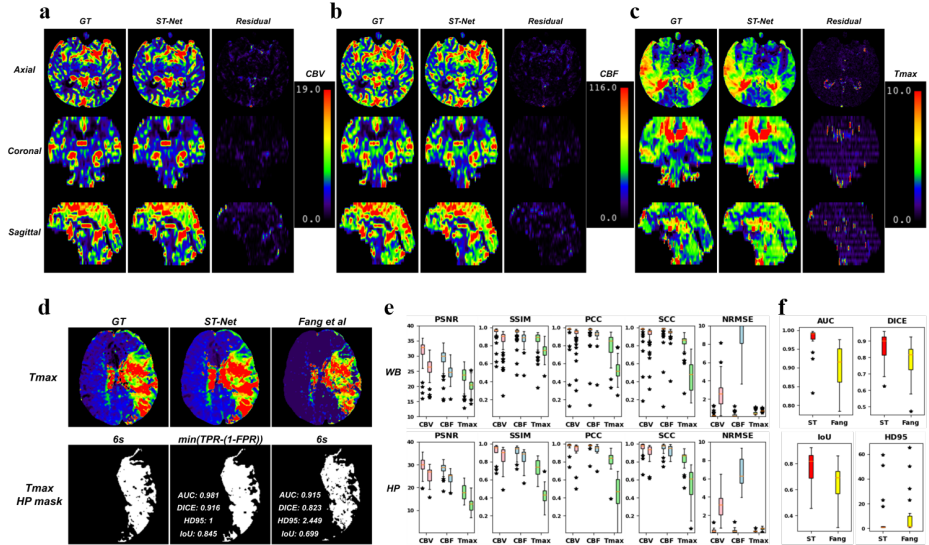

**Fig. 2. a-c.** Perfusion parameter maps obtained from gold standard and the ST-Net, along with the absolute difference (Residual map) between the results from these two methods. CBV: cerebral blood volume, CBF: cerebral blood flow, Tmax: time to maximum. **d.** GT, ST-Net and Fang et al [14] represent Tmax maps derived from the gold standard, ST-Net model and traditional deconvolution method, respectively. HP mask represents hypo-perfused area masks and are derived from Tmax maps. min(TPR-(1-FPR)): The threshold of ROC under this operating point. **e.** Box plots visualizing model performance across testing data. WB: whole brain area, HP: hypo-perfused area. For every parameter per metric, there are two bars with the same color. One is the result of ST-Net, and the other is the deconvolution method (Fang et al [14]). The performance of each metric using the proposed algorithm is consistently significantly better than the traditional deconvolution method, with a P-value less than 0.05. **f.** Box plots visualizing segmentation performance across 18 testing data with lesion labels. Red bar is ST-Net, yellow bar is the Fang et al [14].

### 2.4. Evaluation and statistical analysis

We quantitatively analyzed the similarity between the model's output and the gold standard generated by RAPID. To investigate any advantage of our model, we also displayed the comparison between the deconvolution method (Fang et al [14]) and the gold standard. Several metrics were used, including the Peak Signal-to-Noise Ratio (PSNR) [15], Structural Similarity Index (SSIM) [15], Pearson Correlation Coefficient (PCC) [16], Spearman's Rank Correlation Coefficient (SCC) [17], Normalized Root Mean Squared Error (NRMSE) [18]. Besides the voxel level comparison, we utilized the Area Under the Curve (AUC), Dice Coefficient score (DICE), Intersection over Union (IoU), and Hausdorff Distance at the 95th percentile (HD95) to compare the performance of hypo-perfused region segmentation.

### 3. RESULTS

#### 3.1. Agreement of ST-Net in estimating perfusion parameters

Fig. 2[a-c] shows parameter maps obtained by ST-Net and the ground truth for one testing data. The quantitative evaluation results for all 86 testing data and 18 with lesion labels are summarized in Table 2, presented in the mean (standard deviation) format. ST-Net outperforms the deconvolution method across all metrics, as shown in the box plots in Fig. 2e, demonstrating high consistency with the gold standard.

#### 3.2. Segmentation of hypo-perfused area

We attempted various Tmax threshold values for segmentation by analyzing the relationship between TPR (True Positive Rate) and FPR (False Positive Rate) in the ROC (Receiver Operating Characteristic) curve. We found that the best segmentation performance occurs when the difference between TPR and (1 - FPR) is minimized, with a mean segmentation threshold of 6.484s, outperforming the clinical 6s gold standard. Our method consistently outperformed traditional methods on all metrics, as shown in Table 1 and Fig. 2f. Additionally, Fig. 2d illustrates the segmentation results of one testing data.

**Table 1.** The performance of different methods in segmenting the hypo-perfused areas based on the Tmax parameter in the mean(standard deviation) format. Fang et al is the deconvolution method in [14]. OURS is the ST-Net model. The performance of each metric using the proposed algorithm is significantly better than the traditional method, with a P-value much less than 0.001, except for the HD95 metric, which has a P-value of 0.145.

| DATA | | Hypo-perfused area of 18 testing data with lesion label | | | |
|---|---|---|---|---|---|
| metric | | AUC↑ | DICE↑ | IoU↑ | HD95↓ |
| Tmax | Fang et al | 0.900(0.059) | 0.774(0.116) | 0.645(0.143) | 11.312(18.679) |
| | OURS | 0.974(0.039) | 0.859(0.099) | 0.765(0.141) | 9.268(17.937) |

#### 3.3. Influence of training strategies

**Table 2. Quantitative comparison of parameters obtained from gold standard and derived images.** One main column is in the whole brain area of 86 testing data, the other is in the hypo-perfused area of 18 testing data with lesion labels. Two methods' results are shown. Fang et al represents the deconvolution method in [14]. OURS is the ST-Net. The performance of each metric using the proposed algorithm is consistently significantly better than the deconvolution method, with a P-value less than 0.05.

| DATA | | Whole brain area of all 86 testing data | | | | | Hypo-perfused area of 18 testing data with lesion label | | | | |
|------|------|------|------|------|------|------|------|------|------|------|------|
| metric | | PSNR↑ | SSIM↑ | PCC↑ | SCC↑ | NRMSE↓ | PSNR↑ | SSIM↑ | PCC↑ | SCC↑ | NRMSE↓ |
| CBV | Fang et al | 26.075(3.105) | 0.862(0.116) | 0.925(0.104) | 0.919(0.102) | 2.715(2.067) | 25.202(3.756) | 0.848(0.129) | 0.915(0.133) | 0.894(0.110) | 3.184(1.641) |
|  | OURS | **31.644(3.642)** | **0.952(0.067)** | **0.959(0.117)** | **0.961(0.107)** | **0.252(0.200)** | **29.058(4.600)** | **0.331(0.274)** | **0.965(0.064)** | **0.942(0.086)** | **0.331(0.274)** |
| CBF | Fang et al | 24.470(2.479) | 0.875(0.067) | 0.922(0.067) | 0.941(0.064) | 11.842(4.612) | 23.836(2.924) | 0.835(0.105) | 0.936(0.067) | 0.924(0.064) | 7.685(4.136) |
|  | OURS | **29.125(3.312)** | **0.943(0.073)** | **0.956(0.114)** | **0.966(0.108)** | **0.249(0.170)** | **28.018(3.544)** | **0.909(0.080)** | **0.968(0.055)** | **0.944(0.074)** | **0.355(0.280)** |
| Tmax | Fang et al | 20.354(2.046) | 0.736(0.071) | 0.525(0.094) | 0.456(0.176) | 0.743(0.085) | 12.410(3.220) | 0.426(0.120) | 0.429(0.212) | 0.514(0.239) | 0.547(0.173) |
|  | OURS | **23.366(2.937)** | **0.863(0.094)** | **0.791(0.158)** | **0.828(0.139)** | **0.502(0.193)** | **18.113(3.769)** | **0.723(0.110)** | **0.807(0.125)** | **0.820(0.117)** | **0.288(0.125)** |

We first evaluated the model's performance based on the number of output parameters, concluding that simultaneous output of three parameters (CBV, CBF, and Tmax) significantly enhanced performance. Introducing a physical loss function further improved the model's performance. The minimal difference between the two scenarios also suggests that the original model exhibits sufficient stability. We also analyzed different strides and ratios, finding that decreasing stride improved performance, while the ratio of normal to abnormal voxel sampling had minimal impact, as shown in Fig. 3[a, b]. A 2:1 ratio was chosen for improved training efficiency. Experimenting with learning rates, we found that 1e-4 yielded the most significant effects.

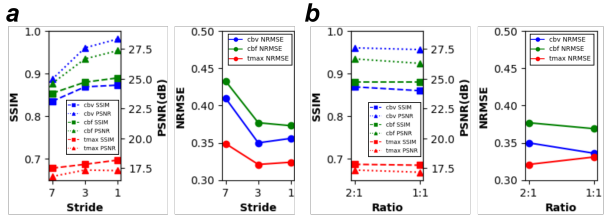

**Fig. 3. a.** The influence of the stride between patches on the model's performance in the hypo-perfused area. **b.** The influence of the ratio of the normal to lesion voxels on the model's performance in the hypo-perfused area.

### 3.4. Computation time

When considering the time required to process 10,000 voxels as a unit, the traditional deconvolution method (Fang et al [14]) took 6.9 seconds. In contrast, the ST-Net model required 0.77 seconds (with a batch size 512). ST-Net is nearly on par with commercial software like RAPID.

### 4. DISCUSSION

ST-Net accurately estimates perfusion parameters from DSC-MRI, showing strong consistency with ground truth. The model's Tmax aligns significantly with the clinical gold standard for identifying hypo-perfused areas, surpassing traditional deconvolution methods like Fang et al [14]. Notably, it's time-efficient and cost-effective. This study innovatively considers both spatial and temporal information, along with an unsupervised loss function—an area largely overlooked in previous research.

While this network model can rapidly and accurately estimate perfusion parameters, it has certain limitations. Firstly, it necessitates preprocessing of the DSC-MRI raw data, including manual delineation of the AIF and VOF. Future research could explore the application of deep learning to automatically extract them from DSC-MRI data. Additionally, data availability constrained this study. Notably, our study revealed that reducing the stride enhances lesion area accuracy, as the stride determines the final number of sampled voxels in the lesion area. Furthermore, disparities in quantification methods indicate that deep learning and traditional methods necessitate distinct strategies for identifying abnormal areas. Prior studies [19, 20] have already demonstrated the potential of deep learning techniques in stroke lesion segmentation, and future research could explore leveraging multi-imaging and multi-parameter information to further enhance the accuracy of identifying abnormal areas following a stroke.

In clinical practice, DSC-MRI requires the injection of contrast agents (GBCAs) to obtain perfusion parameters. However, these agents pose potential hazards [21]. Prior research [22] has shown the feasibility of using deep learning techniques to reduce the amount of contrast agent while maintaining image quality. Developing an alternative framework for imaging without contrast agents, in conjunction with the ST-Net, holds significant clinical significance. AI-based automation of imaging, quantification, and analysis could significantly reduce costs and minimize harm to the human body, which will transform the current clinical treatment landscape.

### 5. CONCLUSION

This study proposes a convolutional neural network that comprehensively considers spatial and temporal information to estimate cerebral perfusion parameters. The ST-Net generates perfusion maps and hypo-perfused area masks, demonstrating high consistency with clinical gold standards, without relying on traditional deconvolution methods. With larger datasets and data parallelism techniques, the model's accuracy and computational speed could be improved, positioning it as an alternative to traditional deconvolution methods for practical quantification of clinical perfusion parameters. This achievement is expected to positively impact clinical practice.

### 6. COMPLIANCE WITH ETHICAL STANDARDS

The studies involving human participants were reviewed and approved by the Ethics Committee of Shanghai Fourth People's Hospital affiliated with the Tongji University School of Medicine,

China (Approval Code: 20200066-01; Approval Date, 15 May 2020). Patient consent was waived due to the nature of the retrospective study.

## 7. ACKNOWLEDGMENTS

This research was funded by the National Key Research and Development Program of China, grant number 2022YFF0710800; the National Key Research and Development Program of China, grant number 2022YFF0710802; the National Natural Science Foundation of China, grant number 62071311; the special program for key fields of colleges and universities in Guangdong Province (biomedicine and health) of China, grant number 2021ZDZX2008; and the Stable Support Plan for Colleges and Universities in Shenzhen of China, grant number SZWD2021010.